\documentclass[conference,usletter]{IEEEtran}
\usepackage{times,amsmath,amssymb}
\usepackage{slashbox}
\usepackage{graphicx}
\usepackage{xcolor}
\usepackage{algorithm2e}
\usepackage{algorithmicx}
\usepackage{setspace}
\usepackage[export]{adjustbox}
\usepackage{multirow}
\usepackage[noend]{algpseudocode}
\usepackage{textcomp}
\newtheorem{defn}{Definition}
\begin{document}

\title{Tensor Representation in High-Frequency Financial Data for Price Change Prediction}
\author{\IEEEauthorblockN{Dat Thanh Tran\IEEEauthorrefmark{1}, Martin Magris\IEEEauthorrefmark{2}, Juho Kanniainen\IEEEauthorrefmark{2}, Moncef Gabbouj\IEEEauthorrefmark{1} \& Alexandros Iosifidis\IEEEauthorrefmark{3}}
\IEEEauthorblockA{\IEEEauthorrefmark{1}Laboratory of Signal Processing, Tampere University of Technology, Tampere, Finland\\
\IEEEauthorrefmark{2}Laboratory of Industrial and Information Management, Tampere University of Technology, Tampere, Finland\\
\IEEEauthorrefmark{3}Department of Engineering, Electrical \& Computer Engineering, Aarhus University, Aarhus, Denmark\\
Email:\{dat.tranthanh,martin.magris,juho.kanniainen,moncef.gabbouj\}@tut.fi, alexandros.iosifidis@eng.au.dk}\\

}

\maketitle

\begin{abstract}
Nowadays, with the availability of massive amount of trade data collected, the dynamics of the financial markets pose both a challenge and an opportunity for high-frequency traders. In order to take advantage of the rapid, subtle movement of assets in High-Frequency Trading (HFT), an automatic algorithm to analyze and detect patterns of price change based on transaction records must be available. The multichannel, time-series representation of financial data naturally suggests tensor-based learning algorithms. In this work, we investigate the effectiveness of two multilinear methods for the mid-price prediction problem against other existing methods. The experiments in a large-scale dataset which contains more than 4 million limit orders show that by utilizing tensor representation, multilinear models outperform vector-based approaches and other competing ones.
\end{abstract}

\section{Introduction}\label{S:Intro}
High-Frequency Trading (HFT) is a form of automated trading that relies on the rapid, subtle changes of the markets to buy or sell assets. The main characteristic of HFT is high speed and short-term investment horizon. Different from long-term investors, high-frequency traders profit from a low margin of the price changes with large volume within a relatively short time. This requires the ability to observe the dynamics of the market to predict prospective changes and act accordingly. In quantitative analysis, mathematical models have been employed to simulate certain aspects of the financial market in order to predict asset price, stock trends, etc. The performance of traditional mathematical models relies heavily on hand-crafted features. With recent advances in computational power, more and more machine learning models have been introduced to predict financial market behaviors. Popular machine learning methods in HFT include regression analysis \cite{zheng2012price,alvim2010daily, pai2005hybrid, detollenaere2017identifying, panayi2016designating}, multilayer feed forward network \cite{levendovszky2012prediction}, \cite{sirignano2016deep}, \cite{galeshchuk2016neural}, convolutional neural network \cite{tsantekidis2017fore} , recurrent neural network \cite{dixon2016high}, \cite{rehman2014foreign}, \cite{tsantekidis2017using}.

With a large volume of data and the erratic behaviors of the market, neural network-based solutions have been widely adopted to learn both the suitable representation of the data and the corresponding classifiers. This resolves the limitation in hand-crafted models. All kinds of deep architectures have been proposed, ranging from traditional multilayer feed-forward models \cite{levendovszky2012prediction}, \cite{sirignano2016deep}, \cite{galeshchuk2016neural} to Convolutional Neural Network (CNN) \cite{tsantekidis2017fore}, Recurrent Neural Network (RNN) \cite{dixon2016high}, \cite{rehman2014foreign}, \cite{tsantekidis2017using}, Deep Belief Networks \cite{sharang2015using}, \cite{hallgren2016testing}, \cite{sandoval2015computational}. For example, in \cite{tsantekidis2017fore} a CNN with both 2D and 1D convolution masks was trained to predict stock price movements. On a similar benchmark HFT dataset, an RNN with Long Short-Term Memory Units (LSTM) \cite{tsantekidis2017using} and or a Neural Bag-of-Features (N-BoF) \cite{passali2017time} network generalizing the (discriminant) Bag-of-Feature model (BoF) \cite{iosifidis2014discriminant} were proposed to perform the same prediction task.

Tensor representation offers a natural representation of the time-series data, where time corresponds to one of the tensor orders. Therefore, it is intuitive to investigate machine learning models that utilize tensor representations. In traditional vector-based models, the features are extracted from the time-series representation and form an input vector to the model. The preprocessing step to convert a tensor representation to a vector representation might lead to the loss of temporal information. That is, the learned classifiers might fail to capture the interactions between spatio-temporal information due to vectorization. Because  many neural network-based solutions, such as CNN or RNN, learn the data directly in the tensor form, this could explain why many neural network implementations outperform traditional vector-based models with hand-crafted features. With advances in mathematical tools and algorithms dealing with tensor input, many multilinear discriminant techniques, as well as tensor regression models have been proposed for image and video classification problems such as \cite{shashua2001linear}, \cite{yang2004two}, \cite{liu1993algebraic}, \cite{kong2005two}, \cite{ye2005two}, \cite{he2006tensor}, \cite{cai2005subspace}, \cite{vasilescu2003multilinear}, \cite{yan2005discriminant}, \cite{tao2007general}. However, there are few works investigating the performance of the tensor-based multilinear methods in financial problems \cite{li2016tensor}. Different from neural network methodology which requires heavy tuning of network topology and parameters, the beauty of tensor-based multilinear techniques is that the objective function is straightforward to interpret and very few parameters are required to tune the model. In this work, we propose to use two multilinear techniques based on the tensor representation of time-series financial data to predict the mid-price movement based on information obtained from Limit Order Book (LOB) data. Specifically, the contribution of this paper is as follows
\begin{itemize}
\item We investigate the effectiveness of tensor-based discriminant techniques, particularly Multilinear Discriminant Analysis (MDA) in a large-scale prediction problem of mid-price movement with high-frequency limit order book data.
\item We propose a simple regression classifier that operates on the tensor representation, utilizing both the current and past information of the stock limit order book to boost the performance of the vector-based regression technique. Based on the observation of the learning dynamics of the proposed algorithm, efficient scheme to select the best model's state is also discussed.
\end{itemize}

The rest of the paper is organized as follows. Section 2 reviews the mid-price movement prediction problem given the information collected from LOB as well as related methods that were proposed to tackle this problem. In Section 3, MDA and our proposed tensor regression scheme are presented. Section 4 shows the experimental analysis of the proposed methods compared with existing results on a large-scale dataset. Finally, conclusions are drawn in Section 5.

\section{High-Frequency Limit Order Data}
In finance, a limit order placed with a bank or a brokerage is a type of trade order to buy or sell a set amount of assets with a specified price. There are two types of limit order: a buy limit order and a sell limit order. In a sell limit order (ask), a minimum sell price and the corresponding volume of assets are specified. For example, a sell limit order of $1000$ shares with a minimum prize of \$$20$ per share indicates that the investors wish to sell the share with maximum prize of \$$20$ only. Similarly, in a buy limit order (bid), a maximum buy price and its respective volume must be specified. The two types of limit orders consequently form two sides of the LOB, the bid and the ask side. LOB aggregates and sorts the order from both sides based on the given price. The best bid price $p^{(1)}_b(t)$ at the time instance $t$ is defined as the highest available price that the buyer is willing to pay per share. The best ask price $p^{(1)}_a(t)$ is, in turn, the lowest available price at a time instance $t$ that a seller is willing to sell per share. The LOB is sorted so that best bid and ask price is on top of the book. The trading happens through a matching mechanism based on several conditions. When the best bid price exceeds the best ask price, i.e. $p^{(1)}_b(t)>p^{(1)}_a(t)$, the trading happens between the two investors. In addition to executions, the order can disappear from the order book by cancellations.

Given the availability of LOB data, several problems can be formulated, such as price trend prediction, order flow distribution estimation or detection of anomalous events that cause turbulence in the price change. One of the popular tasks given the availability of LOB data is to predict the mid-price movements, i.e. to classify whether the mid-price increases, decreases or remains stable based on a set of measurements. The mid-price is a quantity defined as the mean between the best bid price and the best ask price at a given time, i.e.
\begin{equation}
p_t=\frac{p^{(1)}_a(t)+p^{(1)}_b(t)}{2}
\end{equation}
which gives a good estimate of the price trend.

The LOB dataset \cite{ntakaris2017benchmark} used in this paper, referred as FI-2010, was collected from $5$ different Finnish stocks (Kesko, Outokumpu, Sampo, Rautaruukki and Wartsila) in $5$ different industrial sectors. The collection period is from 1st of June to 14th of June 2010, producing order data of $10$ working days. The provided data was extracted based on event inflow \cite{li2016empirical} which aggregates to approximately $4.5$ million events. Each event contains information from the top $10$ orders from each side of the LOB. Since each order consists of a price (bid or ask) and a corresponding volume, each order event is represented by a $40$-dimensional vector. In \cite{ntakaris2017benchmark}, a $144$-dimensional feature vector was extracted for every $10$ events, leading to $453,975$ feature vector samples. For each feature vector, FI-2010 includes an associated label which indicates the movement of mid-price (increasing, decreasing, stationary) in the next $10$ order events. In order to avoid the effect of different scales from each dimension, the data was standardized using z-score normalization
\begin{equation}\
\mathbf{x}_{norm}=\frac{\mathbf{x}-\bar{\mathbf{x}}}{\sigma_{\mathbf{x}}}
\end{equation}

Given the large-scale of FI-2010, many neural network solutions have been proposed to predict the prospective movement of the mid-price. In \cite{tsantekidis2017fore}, a CNN that operates on the raw data was proposed. The network consists of $8$ layers with an input layer of size $100\times 40$, which contains $40$-dimensional vector representation of $100$ consecutive events. The hidden layers contain both 2D and 1D convolution layers as well as max pooling layer. In \cite{tsantekidis2017using}, an RNN architecture with LSTM units that also operates on a similar raw data representation was proposed with separate normalization schemes for order prices and volumes. Beside conventional deep architecture, an N-BoF classifier \cite{passali2017time} was proposed for the problem of the mid-price prediction. The N-BoF network in \cite{passali2017time} was trained on $15$ consecutive $144$-dimensional feature vectors which contain order information from $150$ most recent order events and predicted the movements in the next $k=\{10,50,100\}$ order events. 

It should be noted that all of the above mentioned neural network solutions utilized not only information from the current order events but also information from the recent past. We believe that the information of the recent order events plays a significant role in modeling the dynamics of the mid-price. The next section presents MDA classifier and our proposed regression model that take into account the contribution of past order information.

\section{Tensor-based Multilinear Methods for Financial Data}\label{S:MDA}
Before introducing the classifiers to tackle mid-price prediction problem, we will start with notations and concepts used in multilinear algebra.

\subsection{Multilinear Algebra Concepts}
In this paper, we denote scalar values by either low-case or upper-case characters $(x, y, X, Y \dots)$, vectors by lower-case bold-face characters $(\mathbf{x}, \mathbf{y}, \dots)$, matrices by upper-case bold-face characters $(\mathbf{A}, \mathbf{B}, \dots)$ and tensor as calligraphic capitals $(\mathcal{X}, \mathcal{Y}, \dots)$. A tensor with $K$ modes and dimension $I_{k}$ in the mode-$k$ is represented as $\mathcal{X} \in \mathbb{R}^{I_1 \times I_2 \times \dots \times I_K}$. The entry in the $i_k$th index in mode-$k$ for $k=1,\dots, K$ is denoted as $\mathcal{X}_{i_1,i_2,\dots,i_K}$.

\begin{defn}[Mode-$k$ Fiber and Mode-$k$ Unfolding]\label{def1}
The mode-$k$ fiber of a tensor $\mathcal{X} \in \mathbb{R}^{I_1 \times I_2 \times \dots \times I_K}$ is a vector of $I_k$-dimensional, given by fixing every index but $i_k$. The mode-$k$ unfolding of $\mathcal{X}$, also known as mode-$k$ matricization, transforms the tensor $\mathcal{X}$ to matrix $\mathbf{X}_{(k)}$, which is formed by arranging the mode-$k$ fibers as columns. The shape of $\mathbf{X}_{(k)}$ is $\mathbb{R}^{I_k \times I_{\bar{k}}}$ with $I_{\bar{k}}=\prod_{i=1,i \neq k}^{K} I_i$.
\end{defn}

\begin{defn}[Mode-$k$ Product]\label{def2}
The mode-$k$ product between a tensor $\mathcal{X}=[x_{i_1},\dots , x_{i_K}] \in  \mathbb{R}^{I_1 \times \dots I_K}$ and a matrix $\mathbf{W}\in \mathbb{R}^{J_{k}\times I_k}$ is another tensor of size $I_1\times \dots \times J_{k}\times \dots \times I_K$ and denoted by $\mathcal{X} \times_{k} \mathbf{W}$. The element of $\mathcal{X} \times_{k} \mathbf{W}$ is defined as $[\mathcal{X}\times_{k}\mathbf{W}]_{i_1, \dots , i_{k-1}, j_k, i_{k+1},\dots, i_K}=\sum_{i_k=1}^{I_K}[\mathcal{X}]_{i_1,\dots,i_{k-1},i_k,\dots, i_K}[\mathbf{W}]_{j_k,i_k}$.
\end{defn}

With the definition of mode-$k$ product and mode-$k$ unfolding, the following equation holds
\begin{equation}\label{eq1}
(\mathcal{X}\times_k\mathbf{W}^{T})_{(k)} = \mathbf{W}^{T}\mathbf{X}_{(k)}
\end{equation}
For convenience, we denote $\mathcal{X}\times_1\mathbf{W}_1\times\dots\times_K
\mathbf{W}_K$ by $\mathcal{X} \prod_{k=1}^{K}\times_k\mathbf{W}_k$.

\subsection{Multilinear Discriminant Analysis}
MDA is the extended version of the Linear Discriminant Analysis (LDA) which utilizes the Fisher criterion \cite{welling2005fisher} as the optimal criterion of the learned subspace. Instead of seeking an optimal vector subspace, MDA learns a tensor subspace in which data from different classes are separated by maximizing the interclass distances and minimizing the intraclass distances. The objective function is thus maximizing the ratio between interclass distances and intraclass distances in the projected space. Formally, let us denote the set of $N$ tensor samples as $\mathcal{X}_1,\dots,\mathcal{X}_N \in \mathbb{R}^{I_1\times \dots \times I_K}, i=1,\dots,N$, each with an associated class label $c_i, i=1,\dots,C$. In addition, $\mathcal{X}_{i,j}$ denotes the $j$th sample from class $c_i$ and $n_i$ denotes the number of samples in class $c_i$. The mean tensor of class $c_i$ is calculated as $\mathcal{M}_i=\frac{1}{n_i}\sum_{j=1}^{n_i}\mathcal{X}_{i,j}$ and the total mean tensor is $\mathcal{M}=\frac{1}{N}\sum_{i}^{C}\sum_{j=1}^{n_i}\mathcal{X}_{i,j}=\frac{1}{N}\sum_{i=1}^{C}n_i\mathcal{M}_i$.

MDA seeks a set of projection matrices $\mathbf{W}_k\in \mathbb{R}^{I_k \times I_{k}^{'}}, I_{k}^{'}<I_k, k=1,\dots,K$ that map $\mathcal{X}_{i,j}$ to $\mathcal{Y}_{i,j}\in \mathbb{R}^{I_{1}^{'}\times \dots \times I_{K}^{'}}$, with the subspace projection defined as
\begin{equation}\label{eq6}
 \mathcal{Y}_{i,j}=\mathcal{X}_{i,j}\prod_{k=1}^{K}\times_{k}\mathbf{W}_{k}^{T}
\end{equation}

The set of optimal projection matrices are obtained by maximizing the ratio between interclass and intraclass distances, measured in the tensor subspace $\mathbb{R}^{I_{1}^{'}\times \dots \times I_{K}^{'}}$. Particularly, MDA maximizes the following criterion
\begin{equation}\label{eq7}
J(\mathbf{W}_{1},\dots,\mathbf{W}_K)=\frac{D_b}{D_w}
\end{equation}
where
\begin{equation}\label{eq8}
D_b=\sum_{i=1}^{C}n_i\Vert \mathcal{M}_{i}\prod_{k=1}^{K}\times_{k}\mathbf{W}_k-\mathcal{M}\prod_{k=1}^{K}\times_{k}\mathbf{W}_k\Vert_{F}^{2}
\end{equation}
and
\begin{equation}\label{eq9}
D_w=\sum_{i=1}^{C}\sum_{j=1}^{n_i}\Vert \mathcal{X}_{i,j}\prod_{k=1}^{K}\times_{k}\mathbf{W}_k-\mathcal{M}_{i}\prod_{k=1}^{K}\times_{k}\mathbf{W}_k \Vert_{F}^{2}
\end{equation}
are respectively interclass distance and intraclass distance. The subscript $F$ in (\ref{eq8}) and (\ref{eq9}) denotes the Frobenius norm. $D_b$ measures the total square distances between each class mean $\mathcal{M}_i$ and the global mean $\mathcal{M}$ after the projection while $D_w$ measures the total square distances between each sample and its respective mean tensor. By maximizing (\ref{eq7}), we are seeking a tensor subspace in which the dispersion of data in the same class is minimum while the dispersion between each class is maximum. Subsequently, the classification can then be performed by simply selecting the class with a minimum distance between a test sample to each class mean in the discriminant subspace.
Since the projection in (\ref{eq6}) exposes a dependancy between each mode-$k$, each $\mathbf{W}_k$ cannot be optimized independently. An iterative approach is usually employed to solve the optimization in (\ref{eq7}) [\cite{tao2007general}, \cite{yan2005discriminant}, \cite{li2014multilinear}. In this work, we propose to use the CMDA algorithm \cite{li2014multilinear} that assumes orthogonal constraints on each projection matrix $\mathbf{W}_{k}^{T}\mathbf{W}_k=\mathbf{I}, k=1,\dots,K$ and solves (\ref{eq7}) by iteratively solving a trace ratio problem for each mode-$k$. Specifically, $D_b$ and $D_w$ can be calculated by unfolding the tensors in mode-$k$ as follows
\begin{equation}\label{eq10}
\begin{split}
D_b{} =&tr\bigg(\sum_{i=1}^{C}n_i\Big[\big(\mathcal{M}_i-\mathcal{M}\big)\prod_{p=1}^{K}\times_p \mathbf{W}_{p}^{T}\Big]_{(k)}\\
&\Big[\big(\mathcal{M}_i-\mathcal{M}\big)\prod_{p=1}^{K}\times_p \mathbf{W}_{p}^{T}\Big]_{(k)}^{T}
\bigg)
\end{split}
\end{equation}
and
\begin{equation}\label{eq11}
\begin{split}
D_w{} =&tr\bigg(\sum_{i=1}^{C}\sum_{j=1}^{n_i}\Big[\big(\mathcal{X}_{i,j}-\mathcal{M}_i\big)\prod_{p=1}^{K}\times_p \mathbf{W}_{p}^{T}\Big]_{(k)}\\
&\Big[\big(\mathcal{X}_{i,j}-\mathcal{M}_i\big)\prod_{p=1}^{K}\times_p \mathbf{W}_{p}^{T}\Big]_{(k)}^{T}
\bigg)
\end{split}
\end{equation}
where $tr()$ in (\ref{eq10}) and (\ref{eq11}) denotes the trace operator. By utilizing the identity in (\ref{eq1}), $D_b$ and $D_w$ are further expressed as
\begin{equation}\label{eq12}
\begin{split}
D_b{} &=tr\bigg( \mathbf{W}_{k}^{T}\bigg( \sum_{i=1}^{C} n_i \Big[(\mathcal{M}_i-\mathcal{M})\prod_{p=1, p\neq k}^{K}\times_{p}\mathbf{W}_{p}^{T}\Big]_{(k)} \\
& \Big[(\mathcal{M}_i-\mathcal{M})\prod_{p=1, p\neq k}^{K}\times_{p}\mathbf{W}_{p}^{T}\Big]_{(k)}^{T} \bigg) \mathbf{W}_{k} \bigg)\\
&= tr\big(\mathbf{W}_{k}^{T} \mathbf{S}_{b}^{k} \mathbf{W}_{k}\big)
\end{split}
\end{equation}
and
\begin{equation}\label{eq13}
\begin{split}
D_w{} &= tr\bigg(\mathbf{W}_{k}^{T}\bigg( \sum_{i=1}^{C} \sum_{j=1}^{n_i} \Big[(\mathcal{X}_{i,j}-\mathcal{M}_i)\prod_{p=1, p\neq k}^{K}\times_{p}\mathbf{W}_{p}^{T}\Big]_{(k)} \\
& \Big[(\mathcal{X}_{i,j}-\mathcal{M}_i)\prod_{p=1, p\neq k}^{K}\times_{p}\mathbf{W}_{p}^{T}\Big]_{(k)}^{T} \bigg)  \mathbf{W}_{k} \bigg)\\
&= tr\big(\mathbf{W}_{k}^{T} \mathbf{S}_{w}^{k} \mathbf{W}_{k}\big)
\end{split}
\end{equation}
where $\mathbf{S}_b^k$ and $\mathbf{S}_w^k$ in (\ref{eq12}) and (\ref{eq13}) denote the interclass and intraclass scatter matrices in mode-$k$.  The criterion in (\ref{eq7}) can then be converted to a trace ratio problem with respect to $\mathbf{W}_k$ while keeping other projection matrices fixed as
\begin{equation}\label{eq14}
\mathbf{J}\big(\mathbf{W}_k\big)=\frac{tr\big(\mathbf{W}_{k}^{T} \mathbf{S}_{b}^{k} \mathbf{W}_{k}\big)}{tr\big(\mathbf{W}_{k}^{T} \mathbf{S}_{w}^{k} \mathbf{W}_{k}\big)}
\end{equation}

With the orthogonality constraint of $\mathbf{W}_k$, the solution of (\ref{eq14}) is given by $I_{k}^{'}$ eigenvectors corresponding to $I_{k}^{'}$ largest eigenvalues of $(\mathbf{S}_w^k)^{-1}\mathbf{S}_b$. Usually, a positive $\lambda$ is added to the diagonal of $\mathbf{S}_w^{k}$ as a regularization, which also enables stable computation in case $\mathbf{S}_w^{k}$ is not a full rank matrix. In the training phase, after randomly initializes $\mathbf{W}_k$, CMDA algorithm iteratively goes through each mode $k$, optimizes the Fisher ratio with respect to $\mathbf{W}_k$ while keeping other projection matrices fixed. The algorithm terminates when the changes in $\mathbf{W}_k$ below a threshold or the specified maximum iteration reached. In the test phase, the class with a minimum distance between the class mean and the test sample in the tensor subspace is assigned to the test sample.

\subsection{Weighted Multichannel Time-series Regression}
For the FI-2010 dataset, in order to take into account the past information one could concatenate $T$ $144$-dimensional feature vectors corresponding to the $10T$ most recent order events to form a $2$-mode tensor sample, i.e. a matrix $\mathcal{X}_i \in \mathbb{R}^{144 \times T}, i=1,\dots, N$. For example, a training tensor sample of size $144\times 10$ contains information of $100$ most recent order events in the FI-2010 dataset. $10$ columns represent information at $10$ time-instances with the $10$th column contains the latest order information. Each of the $144$ rows encode the temporal evolution of the $144$ features (or channels) through time. Generally, given $N$ $2$-mode tensor $\mathcal{X}_i \in \mathbb{R}^{D \times T}, i=1,\dots, N$ that belong to $C$ classes indicated by the class label $c_i=1,\dots,C$, the proposed Weighted Multichannel Time-series Regression (WMTR) learns the following mapping function
\begin{equation}\label{eq15}
f\big(\mathcal{X}_i)=\mathbf{W}_{1}^{T}\mathcal{X}_i\mathbf{w}_2
\end{equation}
where $\mathbf{W}_{1} \in \mathbb{R}^{D\times C}$ and $\mathbf{w}_{2} \in \mathbb{R}^{T}$ are learnable parameters. The function $f$ in (\ref{eq15}) maps each input tensor to a $C$-dimensional (target) vector. One way to interpret $f$ is that $\mathbf{W}_1$ maps $D$-dimensional representation of each time-instance to a $C$-dimensional (sub)space while $\mathbf{w}_2$ combines the contribution of each time-instance into a single vector, by using a weighted average approach. In order to deal with unbalanced datasets, such as FI-2010, the parameters $\mathbf{W}_{1}$, $\mathbf{w}_{2}$ of the WMTR model are determined by minimizing the following weighted least square criterion
\begin{equation}\label{eq16}
\begin{split}
J\big(\mathbf{W}_{1},\mathbf{w}_2\big)=& \sum_{i=1}^{N}s_i\Vert \mathbf{W}_{1}^{T}\mathcal{X}_i\mathbf{w}_{2}-\mathbf{y}_i\Vert_{F}^{2}+ \\
&\lambda_1\Vert \mathbf{W}_1\Vert_{F}^{2}+\lambda_2\Vert \mathbf{w}_2\Vert_{F}^2
\end{split}
\end{equation}
where $\mathbf{y}_i\in \mathbb{R}^{C}$ is the corresponding target of the $i$th sample with all elements equal to $-1$ except the $c_i$th element, which is set equal to $1$. $\lambda_1$ and $\lambda_2$ are predefined regularization parameters associated with $\mathbf{W}_1$ and $\mathbf{w}_2$. We set the value of the predefined weight $s_i$ equal to $1 / \sqrt[r]{N_{c_i}}, r >0$, i.e. inversely proportional to the number of training samples belonging to the class of sample $i$, so that errors in smaller classes contribute more to the loss. The weight of each class is controlled by parameter $r$: the smaller $r$, the more contribution of the minor classes in the loss. The unweighted least square criterion is a special case of (\ref{eq16}) when $r \rightarrow +\infty$, i.e. $s_i=1, \forall i$.

We solve (\ref{eq16}) by applying an iterative optimization process that alternatively keeps one parameter fixed while optimizing the other. Specifically, by fixing $\mathbf{w}_2$ we have the following minimization problem
\begin{equation}\label{eq17}
\begin{split}
J_{2}\big(\mathbf{W}_1\big)=& \Vert \big(\mathbf{W}_{1}^{T}\mathbf{X}_{2}-\mathbf{Y}_{2}\big)\mathbf{S}_{2}\Vert_{F}^{2}+ \lambda_1 \Vert \mathbf{W}_1\Vert_{F}^{2}
\end{split}
\end{equation}
where $\mathbf{X}_{2}=\big[\mathcal{X}_1\mathbf{w}_2, \dots, \mathcal{X}_N \mathbf{w}_2\big]\in \mathbb{R}^{D\times N}$, $\mathbf{Y}_{2}=[\mathbf{y}_1,\dots,\mathbf{y}_N] \in \mathbb{R}^{C \times N}$ and $\mathbf{S}_{2}\in \mathbb{R}^{N\times N}$ is a diagonal matrix with the $\mathbf{S}_{2_{i,i}}=\sqrt{s_i}, i=1,\dots,N$. By solving $\frac{\partial J_{2}}{\partial \mathbf{W}_1}=0$, we obtain the solution of (\ref{eq17}) as
\begin{equation}\label{eq18}
\mathbf{W}_{1}^{*}=\big(\mathbf{X}_{2}\mathbf{S}_{2}\mathbf{S}_{2}^{T}\mathbf{X}_{2}^{T}+
\lambda_1\mathbf{I}\big)^{-1}\mathbf{X}_{2}\mathbf{S}_{2}\mathbf{S}_{2}^{T}\mathbf{Y}_{2}^{T}
\end{equation}
where $\mathbf{I}$ is the identity matrix of the appropriate size.

Similarly, by fixing $\mathbf{W}_1$, we have the following regression problem with respect to $\mathbf{w}_2$
\begin{equation}\label{eq19}
J_{1}(\mathbf{w}_2)=\Vert \mathbf{S}_1 \big(\mathbf{X}_{1}\mathbf{w}_{2}-\mathbf{Y}_{1}\big)\Vert_{F}^{2}+ \lambda_2 \Vert \mathbf{w}_{2}\Vert_{F}^{2}
\end{equation}
where $\mathbf{X}_{1}=\big[\mathcal{X}_{1}^T \mathbf{W}_{1}, \dots,\mathcal{X}_{N}^T \mathbf{W}_{1}\big]^{T} \in \mathbb{R}^{CN \times T}$, $\mathbf{Y}_{(1)}=[\mathbf{y}_{1}^T,\dots,\mathbf{y}_{N}^{T}]^{T} \in \mathbb{R}^{CN}$ and $\mathbf{S}_1 \in \mathbb{R}^{CN \times CN}$ is a diagonal matrix with $\mathbf{S}_{1_{C(i-1)+k,C(i-1)+k}}=\sqrt{s_i}; k=1,\dots,C; i=1,\dots,N$. Similar to $\mathbf{W}_1$, optimal $\mathbf{w}_2$ is obtained by solving for the stationary point of (\ref{eq19}), which is given as
\begin{equation}\label{eq20}
\mathbf{w}_{2}^{*}=\big(\mathbf{X}_{1}^{T}\mathbf{S}_{1}^{T}\mathbf{S}_{1}\mathbf{X}_{1}+\lambda_2 \mathbf{I}\big)^{-1}\mathbf{X}_{1}^{T}\mathbf{S}_{1}^{T}\mathbf{S}_{1}\mathbf{Y}_{1}
\end{equation}

The above process is formed by two convex problems, for which each processing step obtains the global optimum solution. Thus, the overall process is guaranteed to reach a local optimum for the combined regression criterion. The algorithm terminates when the changes in $\mathbf{W}_1$ and $\mathbf{w}_2$ are below a threshold, or the maximum number of iterations is reached. In the test phase, $f$ in (\ref{eq15}) maps a test sample to the feature space, and the class label is inferred by the index of the maximum element of the projected test sample.

Usually, multilinear methods (including multilinear regression ones) are randomly initialized. This means that, in our case, one would randomly initialize the parameters in $\mathbf{w}_2$ in order to define the optimal regression values stored in $\mathbf{W}_1$ on the first iteration. However, since for WMTR when applied to LOB data, the values of $\mathbf{w}_2$ encode the contribution of each time-instance in the overall regression, we chose to initialize it as $\mathbf{w}_2 = [0 \:0 \dots 1]^T$. That is, the first iteration of WMTR corresponds to the vector-based regression using only the representation for the current time-instance. After obtaining this mapping, the optimal weighted average of all time-instances is determined by solving for $\mathbf{w}_2$. 

\section{Experiments}\label{S:Experiments}
\subsection{Experiment Setting}
\begin{figure}[t!]
\centering
\includegraphics[width=.5\textwidth]{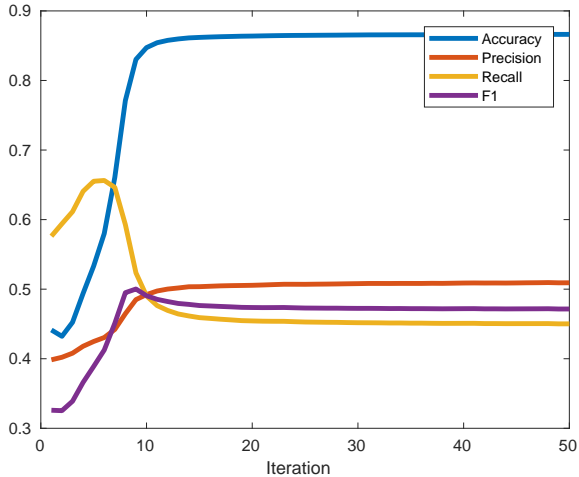}
\caption{Performance measure of WMTR on training data\label{overflow}}
\end{figure}

We conducted extensive experiments on the FI-2010 dataset to compare the performance of the multilinear methods, i.e. MDA and the proposed WMTR, with that of the other existing methods including LDA, Ridge Regression (RR), Single-hidden Layer Feed Forward Network (SLFN), BoF and N-BoF. In addition, we also compared WMTR with its unweighted version, denoted by MTR, to illustrate the effect of weighting in the learned function. Regarding the train/test evaluation protocol, we followed the anchored forward cross-validation splits provided by the database \cite{ntakaris2017benchmark}. Specifically, there are 9 folds for cross-validation based on the day basis; the training set increases by one day for each consecutive fold and the day following the last day used for training is used for testing, i.e. for the first fold, data from the first day is used for training and data from the second day is used for testing; for the second fold, data from the first and second day is used as for training and data from the third day used for testing; for the last fold, data from the first 9 days is used for training and the $10$th day is used for testing.

Regarding the input representation of the proposed multilinear techniques, MDA and WMTR both accept input tensor of size $\mathbb{R}^{144\times 10}$, which contains information from $100$ consecutive order events with the last column contains information from the last $10$ order events. For LDA, RR and SLFN, each input vector is of size $\mathbb{R}^{144}$, which is the last column of the input from MDA and WMTR, representing the most current information of the stock. The label of both tensor input and vector input is the movement of the mid-price in the next $10$ order events, representing the future movement that we would like to predict. Since we followed the same experimental protocol as in \cite{ntakaris2017benchmark} and \cite{passali2017time}, we directly report the result of RR, SLFN, BoF, N-BoF in this paper.

The parameter settings of each model are as follows. For WMTR, we set maximum number of iterations to $50$, the terminating threshold to $1e-6$; $\lambda_1, \lambda_2 \in \{0.01,0.1,1,10,100\}$ and $s_i=n_{c_i}^{-1/r}$ with $r \in \{2,3,4\}$.
For MTR, all paramter settings were similar to WMTR except $s_i=1,\: \forall i$. For MDA, the number of maximum iterations and terminating threshold were set similar to WMTR, the projected dimensions of the first mode is from $5$ to $60$ with a step of $5$ while for the second mode from $1$ to $8$ with a step of $1$. In addition, a regularization amount $\lambda \in \{0.01, 0.1, 1, 10, 100\}$ was added to the diagonal of $\mathbf{S}_{w}^{k}$.
\begin{figure}[t!]
\centering
\includegraphics[width=.5\textwidth]{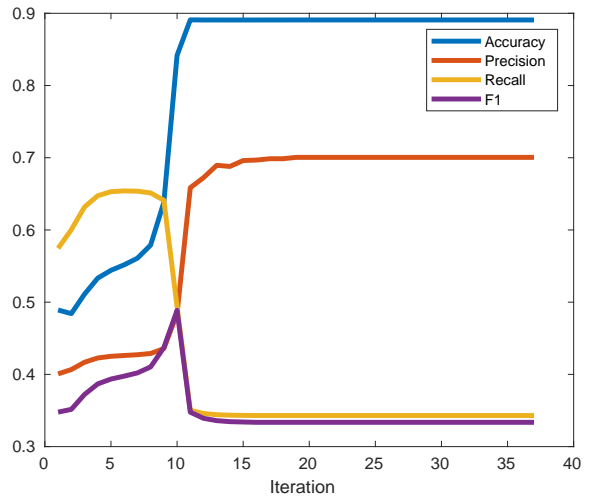}
\caption{Performance measure of MTR on training data\label{overflow}}
\end{figure}

\subsection{Performance Evaluation}

\begin{table}[b!]\label{t1}
\begin{center}
\caption{Performance on FI-2010}
\resizebox{\linewidth}{!}{
\begin{tabular}{|l|c|c|c|c|}\cline{2-5}
\multicolumn{1}{c|}{}
		& Accuracy 			& Precision			& Recall			& F1 		\\ \hline
RR		& $46.00\pm 2.85$	& $43.30\pm 9.9$	& $43.54\pm 5.2$	& $42.52\pm 1.22$	\\ \hline		
SLFN	& $53.22\pm 7.04$	& $49.60\pm 3.81$	& $41.28\pm 4.04$	& $38.24\pm 5.66$	\\ \hline
LDA		& $63.82\pm 4.98$	& $37.93\pm 6.00$	&$45.80\pm 4.07$	& $36.28\pm 1.02$	\\ \hline
MDA		& $71.92\pm 5.46$	& $44.21\pm 1.35$	&$60.07\pm 2.10$	& $46.06\pm 2.22$	\\ \hline
MTR		& $86.08\pm 4.99$	& $51.68\pm 7.54$	&$40.81\pm 6.18$	& $40.14\pm 5.26$	\\ \hline	
WMTR	& $81.89\pm 3.65$	& $46.25\pm 1.90$	&$51.29\pm 1.88$	& $\mathbf{47.87}\pm \mathbf{1.91}$	\\ \hline
BoF		& $57.59\pm 7.34$	& $39.26\pm 0.94$	&$51.44\pm 2.53$	& $36.28\pm 2.85$	\\ \hline
N-BoF	& $62.70\pm 6.73$	& $42.28\pm 0.87$	&$61.41\pm 3.68$	& $41.63\pm 1.90$	\\ \hline

\end{tabular}
}
\end{center}
\end{table}

It should be noted that FI-2010 is a highly unbalanced dataset with most samples having a stationary mid-price. Therefore we use average $f1$ score per class \cite{powers2011evaluation} as a performance measure to select model parameters since $f1$ expresses a trade-off between precision and recall. More specifically, for each cross-validation fold, the competing methods are trained with all combinations of the above-mentioned parameter settings on the training data. We selected the learned model that achieved the highest $f1$ score on the training set and reported the performance on the test set. In addition to $f1$, the corresponding average precision per class, average recall per class and accuracy are also reported. Accuracy measures the percentage the predicted labels that match the ground truth. Precision is the ratio between true positives over the number of samples predicted as positive, and recall is the ratio between true positive over the total number of true positives and false negatives. $f1$ is the harmonic mean between precision and recall. For all measures, higher values indicate better performance.

\begin{figure}[t!]
\centering
\includegraphics[width=.5\textwidth]{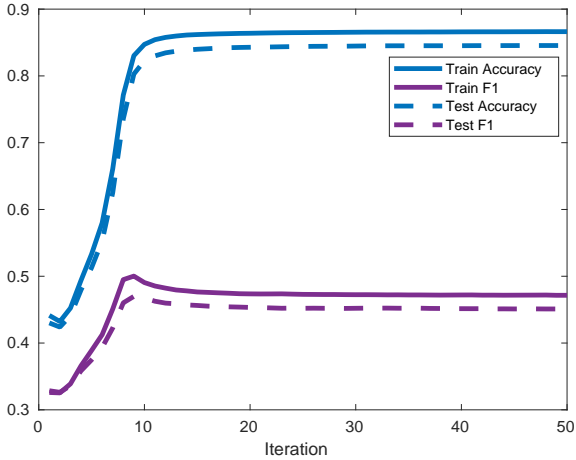}
\caption{Performance measure of WMTR on both train and test set\label{overflow}}
\end{figure}

Table 1 shows the average performance with standard deviation over all 9 folds of the competing methods. Comparing two discriminant methods, i.e. LDA and MDA, it is clear that MDA significantly outperforms LDA on all performance measures. This is due to the fact that MDA operates on the tensor input, which could hold both current and past information as well as the temporal structure of the data. The improvement of tensor-based approaches over vector-based approach is consistent also in case of regression (WMTR vs RR). Comparing multilinear techniques with N-BoF, MDA and WMTR perform much better than N-BoF in terms of $f1$, accuracy and precision while recall scores nearly match. WMTR outperforming MTR in large margin suggests that weighting is important for the highly unbalanced dataset such as FI-2010. Overall, MDA and WMTR are the leading methods among the competing methods in this mid-price prediction problem.

\begin{figure}[t!]
\centering
\includegraphics[width=.5\textwidth]{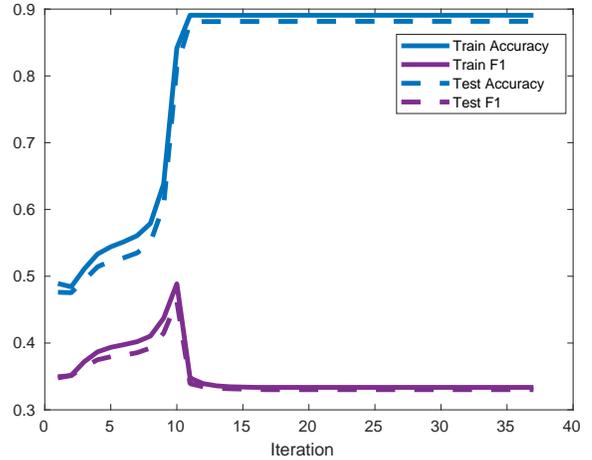}
\caption{Performance measure of MTR on both train and test set\label{overflow}}
\end{figure}

\subsection{WMTR analysis}
Figure 1 shows the dynamic of the learning process of WMTR on the training data of the first fold. There is one interesting phenomenon that can be observed during the training process. In the first $10$ iterations, all performance measures improve consistently. After the $10$th iteration, $f1$ score drops a little then remains stable while accuracy continues to improve. This phenomenon can be observed in every parameter setting. Since WMTR minimizes the squared error between the target label and the predicted label, constant improvement before converging observed from the training accuracy is expected. The drop in $f1$ score after some $k$ iterations can be explained as follows: in the first $k$ iterations, WMTR truly learns the generating process behind the training samples; however, at a certain point, WMTR starts to overfit the data and becomes bias towards the dominant class. The same phenomenon was observed from MTR with a more significant drop in $f1$ since without weight MTR overfits the dominant class severely. Figure 2 shows the training dynamic of MTR with similar parameter setting except for the class weight in the loss function. Due to this behavior, in order to select the best learned state of WMTR and MTR, we measured $f1$ score on the training data at each iteration and selected the model's state which produced the best $f1$. The question is whether the selected model performs well on the test data? Figure 3 and Figure 4 plots accuracy and $f1$ of WMTR and MTR measured on the training set and the test set at each iteration. It is clear that the learned model that produced best $f1$ during training also performed best on the test data. The margin between training and testing performance is relatively small for both WMTR and MTR which shows that our proposed algorithm did not suffer from overfitting. Although the behaviors of WMTR and MTR are similar, the best model learned from MTR is biased towards the dominant class, resulting in inferior performance as shown in the experimental result.

\section{Conclusions}\label{S:Conclusions}
In this work, we have investigated the effectiveness of multilinear discriminant analysis in dealing with financial data prediction based on Limit Order Book data. In addition, we proposed a simple bilinear regression algorithm that utilizes both current and past information of a stock to boost the performance of traditional vector-based regression. Experimental results showed that the proposed methods outperform their counterpart exploiting vectorial representations, and outperform existing solutions utilizing (possibly deep) neural network architectures.

\section{Acknowledgement}
This project has received funding from the European Union\textquotesingle s Horizon 2020 research and innovation programme under the Marie Skłodowska-Curie grant agreement No 675044 “BigDataFinance”.
\bibliography{finance}

\begin{thebibliography}{10}

\bibitem{zheng2012price}
B.~Zheng, E.~Moulines, and F.~Abergel, ``Price jump prediction in limit order
  book,'' 2012.

\bibitem{alvim2010daily}
L.~G. Alvim, C.~N. dos Santos, and R.~L. Milidiu, ``Daily volume forecasting
  using high frequency predictors,'' in {\em Proceedings of the 10th IASTED
  International Conference}, vol.~674, p.~248, 2010.

\bibitem{pai2005hybrid}
P.-F. Pai and C.-S. Lin, ``A hybrid arima and support vector machines model in
  stock price forecasting,'' {\em Omega}, vol.~33, no.~6, pp.~497--505, 2005.

\bibitem{detollenaere2017identifying}
B.~Detollenaere and C.~D'hondt, ``Identifying expensive trades by monitoring
  the limit order book,'' {\em Journal of Forecasting}, vol.~36, no.~3,
  pp.~273--290, 2017.

\bibitem{panayi2016designating}
E.~Panayi, G.~W. Peters, J.~Danielsson, and J.-P. Zigrand, ``Designating market
  maker behaviour in limit order book markets,'' {\em Econometrics and
  Statistics}, 2016.

\bibitem{levendovszky2012prediction}
J.~Levendovszky and F.~Kia, ``Prediction based-high frequency trading on
  financial time series,'' {\em Periodica Polytechnica. Electrical Engineering
  and Computer Science}, vol.~56, no.~1, p.~29, 2012.

\bibitem{sirignano2016deep}
J.~Sirignano, ``Deep learning for limit order books,'' 2016.

\bibitem{galeshchuk2016neural}
S.~Galeshchuk, ``Neural networks performance in exchange rate prediction,''
  {\em Neurocomputing}, vol.~172, pp.~446--452, 2016.

\bibitem{tsantekidis2017fore}
A.~Tsantekidis, N.~Passalis, A.~Tefas, J.~Kanniainen, M.~Gabbouj, and
  A.~Iosifidis, ``Forecasting stock prices from the limit order book using
  convolutional neural networks,'' in {\em IEEE Conference on Business
  Informatics (CBI), Thessaloniki, Greece}, 2017.

\bibitem{dixon2016high}
M.~Dixon, ``High frequency market making with machine learning,'' 2016.

\bibitem{rehman2014foreign}
M.~Rehman, G.~M. Khan, and S.~A. Mahmud, ``Foreign currency exchange rates
  prediction using cgp and recurrent neural network,'' {\em IERI Procedia},
  vol.~10, pp.~239--244, 2014.

\bibitem{tsantekidis2017using}
A.~Tsantekidis, N.~Passalis, A.~Tefas, J.~Kanniainen, M.~Gabbouj, and
  A.~Iosifidis, ``Using deep learning to detect price change indications in
  financial markets,'' in {\em European Signal Processing Conference (EUSIPCO),
  Kos, Greece}, 2017.

\bibitem{sharang2015using}
A.~Sharang and C.~Rao, ``Using machine learning for medium frequency derivative
  portfolio trading,'' {\em arXiv preprint arXiv:1512.06228}, 2015.

\bibitem{hallgren2016testing}
J.~Hallgren and T.~Koski, ``Testing for causality in continuous time bayesian
  network models of high-frequency data,'' {\em arXiv preprint
  arXiv:1601.06651}, 2016.

\bibitem{sandoval2015computational}
J.~Sandoval and G.~Hern{\'a}ndez, ``Computational visual analysis of the order
  book dynamics for creating high-frequency foreign exchange trading
  strategies,'' {\em Procedia Computer Science}, vol.~51, pp.~1593--1602, 2015.

\bibitem{passali2017time}
N.~Passalis, A.~Tsantekidis, A.~Tefas, J.~Kanniainen, M.~Gabbouj, and
  A.~Iosifidis, ``Time-series classification using neural bag-of-features,'' in
  {\em European Signal Processing Conference (EUSIPCO), Kos, Greece}, 2017.

\bibitem{iosifidis2014discriminant}
A.~Iosifidis, A.~Tefas, and I.~Pitas, ``Discriminant bag of words based
  representation for human action recognition,'' {\em Pattern Recognition
  Letters}, vol.~49, pp.~185--192, 2014.

\bibitem{shashua2001linear}
A.~Shashua and A.~Levin, ``Linear image coding for regression and
  classification using the tensor-rank principle,'' in {\em Computer Vision and
  Pattern Recognition, 2001. CVPR 2001. Proceedings of the 2001 IEEE Computer
  Society Conference on}, vol.~1, pp.~I--I, IEEE, 2001.

\bibitem{yang2004two}
J.~Yang, D.~Zhang, A.~F. Frangi, and J.-y. Yang, ``Two-dimensional pca: a new
  approach to appearance-based face representation and recognition,'' {\em IEEE
  transactions on pattern analysis and machine intelligence}, vol.~26, no.~1,
  pp.~131--137, 2004.

\bibitem{liu1993algebraic}
K.~Liu, Y.-Q. Cheng, and J.-Y. Yang, ``Algebraic feature extraction for image
  recognition based on an optimal discriminant criterion,'' {\em Pattern
  Recognition}, vol.~26, no.~6, pp.~903--911, 1993.

\bibitem{kong2005two}
H.~Kong, E.~K. Teoh, J.~G. Wang, and R.~Venkateswarlu, ``Two-dimensional fisher
  discriminant analysis: forget about small sample size problem [face
  recognition applications],'' in {\em Acoustics, Speech, and Signal
  Processing, 2005. Proceedings.(ICASSP'05). IEEE International Conference on},
  vol.~2, pp.~ii--761, IEEE, 2005.

\bibitem{ye2005two}
J.~Ye, R.~Janardan, and Q.~Li, ``Two-dimensional linear discriminant
  analysis,'' in {\em Advances in neural information processing systems},
  pp.~1569--1576, 2005.

\bibitem{he2006tensor}
X.~He, D.~Cai, and P.~Niyogi, ``Tensor subspace analysis,'' in {\em Advances in
  neural information processing systems}, pp.~499--506, 2006.

\bibitem{cai2005subspace}
D.~Cai, X.~He, and J.~Han, ``Subspace learning based on tensor analysis,''
  tech. rep., 2005.

\bibitem{vasilescu2003multilinear}
M.~A.~O. Vasilescu and D.~Terzopoulos, ``Multilinear subspace analysis of image
  ensembles,'' in {\em Computer Vision and Pattern Recognition, 2003.
  Proceedings. 2003 IEEE Computer Society Conference on}, vol.~2, pp.~II--93,
  IEEE, 2003.

\bibitem{yan2005discriminant}
S.~Yan, D.~Xu, Q.~Yang, L.~Zhang, X.~Tang, and H.-J. Zhang, ``Discriminant
  analysis with tensor representation,'' in {\em Computer Vision and Pattern
  Recognition, 2005. CVPR 2005. IEEE Computer Society Conference on}, vol.~1,
  pp.~526--532, IEEE, 2005.

\bibitem{tao2007general}
D.~Tao, X.~Li, X.~Wu, and S.~J. Maybank, ``General tensor discriminant analysis
  and gabor features for gait recognition,'' {\em IEEE Transactions on Pattern
  Analysis and Machine Intelligence}, vol.~29, no.~10, 2007.

\bibitem{li2016tensor}
Q.~Li, Y.~Chen, L.~L. Jiang, P.~Li, and H.~Chen, ``A tensor-based information
  framework for predicting the stock market,'' {\em ACM Transactions on
  Information Systems (TOIS)}, vol.~34, no.~2, p.~11, 2016.

\bibitem{ntakaris2017benchmark}
A.~Ntakaris, M.~Magris, J.~Kanniainen, M.~Gabbouj, and A.~Iosifidis,
  ``Benchmark dataset for mid-price prediction of limit order book data,'' {\em
  arXiv preprint arXiv:1705.03233}, 2017.

\bibitem{li2016empirical}
X.~Li, H.~Xie, R.~Wang, Y.~Cai, J.~Cao, F.~Wang, H.~Min, and X.~Deng,
  ``Empirical analysis: stock market prediction via extreme learning machine,''
  {\em Neural Computing and Applications}, vol.~27, no.~1, pp.~67--78, 2016.

\bibitem{welling2005fisher}
M.~Welling, ``Fisher linear discriminant analysis,'' {\em Department of
  Computer Science, University of Toronto}, vol.~3, no.~1, 2005.

\bibitem{li2014multilinear}
Q.~Li and D.~Schonfeld, ``Multilinear discriminant analysis for higher-order
  tensor data classification,'' {\em IEEE transactions on pattern analysis and
  machine intelligence}, vol.~36, no.~12, pp.~2524--2537, 2014.

\bibitem{powers2011evaluation}
D.~M. Powers, ``Evaluation: from precision, recall and f-measure to roc,
  informedness, markedness and correlation,'' 2011.

\end{thebibliography}
\bibliographystyle{ieeetr}

\end{document}